\documentclass[aps, prl, twocolumn, amsmath, amssymb, showpacs, superscriptaddress, notitlepage, longbibliography]{revtex4-2}
\usepackage{bm}
\usepackage{graphicx}
\usepackage{color}
\usepackage{multirow}
\usepackage{amsmath}
\usepackage{physics}
\usepackage{float}
\usepackage[colorlinks=true, linkcolor=blue, anchorcolor=red, citecolor=blue, urlcolor=blue]{hyperref}

\begin{document}

\title{Quantum oscillation in Hopf-link semimetals}

\author{Lei Shi$^\dag$}
\affiliation{State Key Laboratory of Quantum Functional Materials, Department of Physics,
and Guangdong Basic Research Center of Excellence for Quantum Science, Southern University of Science and Technology (SUSTech), Shenzhen 518055, China}
\affiliation{Quantum Science Center of Guangdong-Hong Kong-Macao Greater Bay Area (Guangdong), Shenzhen 518045, China}

\author{Xiaoxiong Liu$^\dag$}
\affiliation{State Key Laboratory of Quantum Functional Materials, Department of Physics,
and Guangdong Basic Research Center of Excellence for Quantum Science, Southern University of Science and Technology (SUSTech), Shenzhen 518055, China}
\affiliation{Quantum Science Center of Guangdong-Hong Kong-Macao Greater Bay Area (Guangdong), Shenzhen 518045, China}

\author{C. M. Wang$^\dag$}
\affiliation{Department of Physics, Shanghai Normal University, Shanghai 200234, China}

\author{Tianyu Liu}
\affiliation{International Quantum Academy, Shenzhen 518048, China}
\affiliation{Shenzhen Key Laboratory of Quantum Science and Engineering, Shenzhen 518055, China}

\author{Hai-Zhou Lu}
\email{Corresponding author: luhz@sustech.edu.cn}
\affiliation{State Key Laboratory of Quantum Functional Materials, Department of Physics,
and Guangdong Basic Research Center of Excellence for Quantum Science, Southern University of Science and Technology (SUSTech), Shenzhen 518055, China}

\affiliation{Quantum Science Center of Guangdong-Hong Kong-Macao Greater Bay Area (Guangdong), Shenzhen 518045, China}

\author{X. C. Xie}
\affiliation{International Center for Quantum Materials, School of Physics, Peking University, Beijing100871, China}
\affiliation{Institute for Nanoelectronic Devices and Quantum Computing, Fudan University, Shanghai 200433, China}
\affiliation{Hefei National Laboratory, Hefei 230088, China}

\date{\today }

\begin{abstract}
Since the discovery of the relation between the Chern number and quantum Hall effect, searching for observables of topological invariants has been an intriguing topic. Topological Hopf-link semimetals have attracted tremendous interest, in which the conduction and valence energy bands touch at linked nodal lines. However, it is challenging to identify this sophisticated topology. We propose to use the quantum oscillation in strong magnetic fields to probe the Hopf links. For a generic model of Hopf-link semimetal that captures the linked-trivial phase transition, we figure out the phase shifts of oscillation for all Fermi pockets in all magnetic-field directions, by presenting self-consistent results from the Fermi surface tomography, Landau fan diagram, and electrical resistivity. As the magnetic field is rotated, the phase shifts exhibit a unique pattern, which could help to identify Hopf links in real materials, such as those in Li$_2$NaN. 
\end{abstract}

\maketitle

\begin{figure}[h!]
\centering
\includegraphics[width=0.46\textwidth]{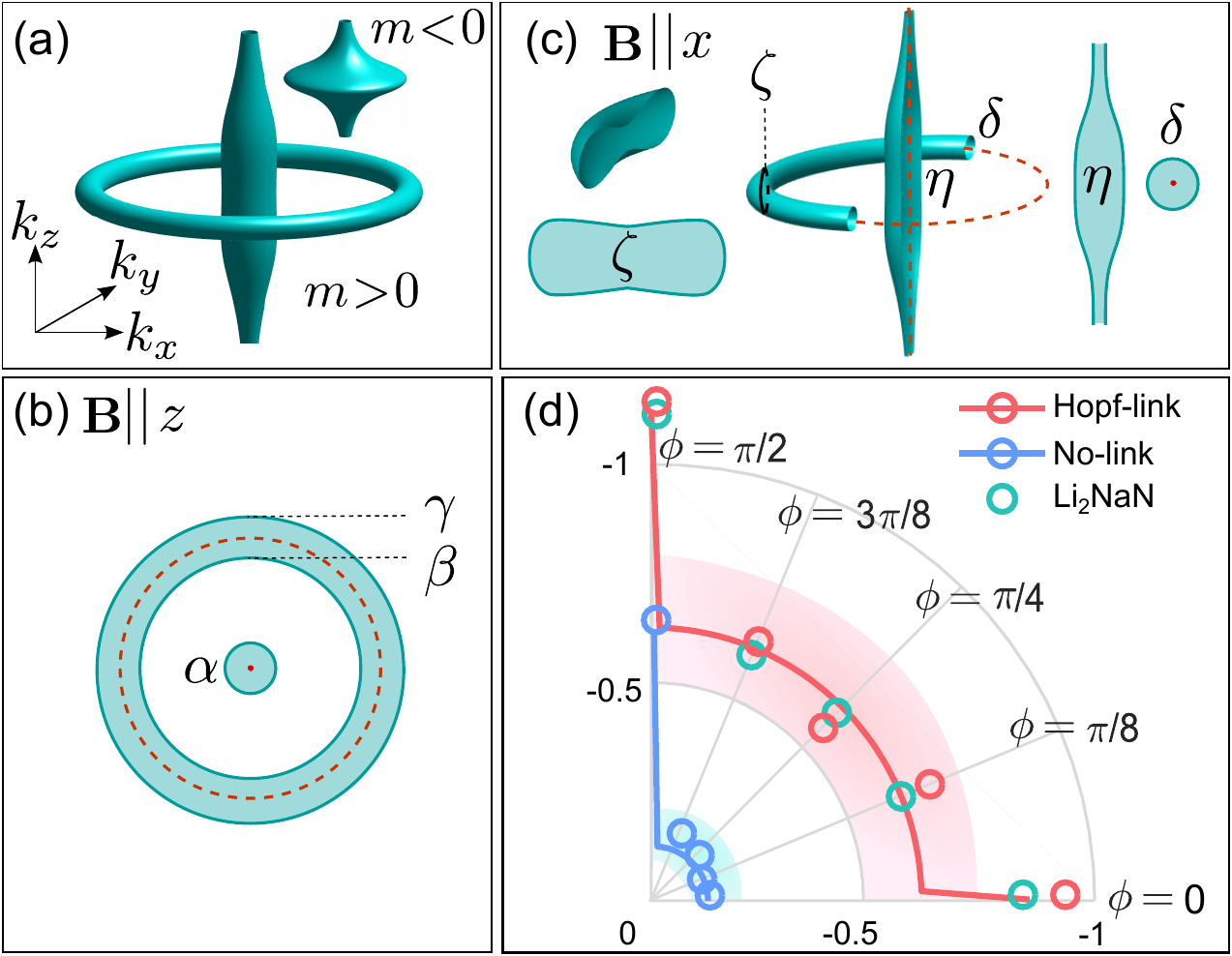}
\caption{(a) Fermi surfaces of a Hopf-link semimetal ($m>0$) and the trivial case ($m<0$). $m$ is a model parameter in Eq. \eqref{Eq:Hamiltonian}. $k_x,k_y,k_z$ are the wave vectors.
[(b) and (c)] Extremal cross sections of the Hopf-link semimetal in a magnetic field $\mathbf{B}$ along the $z$ ($\phi=0$) and $x$ ($\phi=\pi/2$)
directions, respectively. 
The red dashed lines indicate the linked nodal ring and nodal line where the conduction and valence energy bands touch. 
(d) The summation of the phase shifts from all cross sections as a function of $\phi$ in radial coordinates (solid lines from analytical and circles from numerical results) forms a unique pattern for the Hopf-link semimetal and Li$_2$NaN (red shadow), as the three extremal cross sections transit from ($\alpha$, $\beta$, $\gamma$) to ($\eta$, $\delta$, $\zeta$), compared to the no-link case (blue circles) with only one extremal cross section.}
\vspace{-2.5em}
\label{Fig:Model}
\end{figure}

\textcolor{blue}{Introduction} - 
The observable associated with topological invariants is an intriguing topic.
The well-known paradigms include the Chern number for the quantum Hall effects \cite{Klitzing1980PRL, Thouless1982PRL, Lu10prb,Yu2010Science, Chang2013science}, $\mathbb{Z}_2$ invariant for the quantum spin Hall effect \cite{Kane2005PRL}, winding number for the Su-Shrieffer-Heeger model of polyacetylene \cite{Ian2014PRL, SongJT2014PRB}, Euler characteristic for ballistic conductor \cite{Kane2022PRL, Kane2023PRL}, Hopf invariant that maps 3D spheres to 2D \cite{Moore2008PRL, DengDL2013PRB, LiuCX2017hopfPRB, YanZB2023PRB}, skyrmion number for magnetic vortices \cite{Nagaosa2004JPSJ, Nagaosa2013NatureNano}.
Recently, an emerging Hopf-link semimetal has attracted great attention \cite{ChenW2017PRB, YanZB2017PRB, Ezawa17prbrc, Zhong17nc, ZhouY2018PRB}, in which
the conduction and valence energy bands touch at linked nodal lines or nodal rings (Fig.~\ref{Fig:Model}). 
The Hopf links have been realized mainly in bosonic systems, such as photonic crystals \cite{GaoZ2022Nature}, superconducting circuits \cite{TanXS2018APL}, trapped ion simulator \cite{CaoMM2023PRL}, topological Raman lattice \cite{YuJL2019PRA, YiCR2019Arxiv}.
However, in real materials, due to lack of practical probes and clear signatures, the Hopf link has only been observed in Co$_2$MnGa by using angle-resolved photoemission spectroscopy \cite {Hasan2022Nature}.

\begin{figure}[tbp]
\centering
\includegraphics[width=0.46\textwidth]{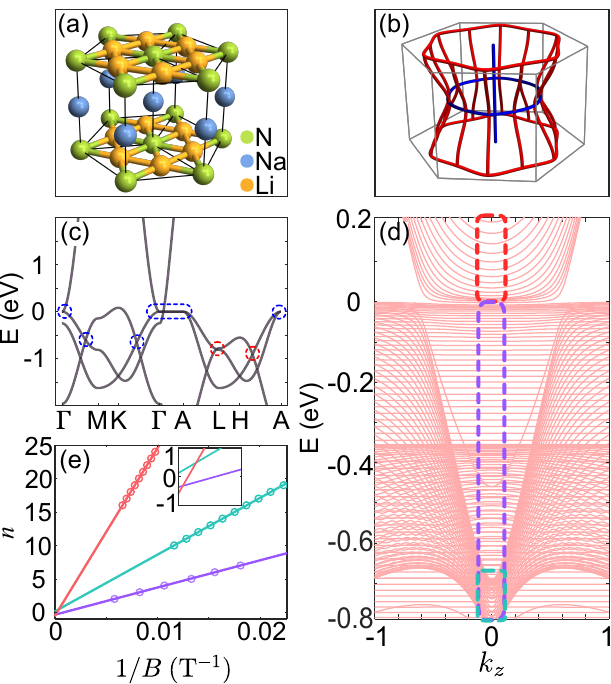}
\caption{The predicted Hopf-link semimetal material Li$_2$NaN \cite{Lenggenhager21prbl}. 
(a) Crystal structure of Li$_2$NaN.
(b) Nodal lines of Li$_2$NaN in the first Brillouin zone, including
a Hopf link (blue) consisting of a nodal-line and a nodal-ring, and additional nodal lines (red).
(c) Band structure described by the tight-binding model in Eq.~(\ref{Eq:LiNaN}) on high-symmetry lines, where Fermi energy is 0 eV. The blue dashed rings indicates the nodal-lines (at Fermi energy) and nodal-rings (-0.6 eV below Fermi energy) of the blue Hopf-link. The red dashed rings indicate the red additional nodal lines.
(d) The Landau-band energies under $B_z$ = 10 T as functions of the wave vector $k_{z}$.
(e) The $n$-$1/B$ plots for three distinct frequencies, extracted from the band-bottom energies (indicated by the dashed boxes at $k_z=0$ in (d)) at the Fermi energy $E_F=-0.75$ (green), -0.2 (purple), and 0.75 eV (red), respectively. The intercepts on the $n$ axis (Inset: Zoom-in) give the phase shifts, as listed in Tab.~\ref{Tab:Li2NaN} at $\phi = 0$. }\label{Fig:Li2NaN}
\end{figure}

In this Letter, we propose that the Hopf links could be probed by the quantum oscillation of resistance, which arises from the Landau quantization of electronic
states under strong magnetic fields. 
The oscillation can
be described by $\cos{[2\pi(F/B+\Phi)]}$, where $B$ is the strength of magnetic field and the oscillation frequency $F$ and phase shift $\Phi$ can provide valuable Fermi-surface and topological properties of energy bands \cite{lifshitz1956theory,Shoenberg84book}. 
We systematically study the phase shifts of quantum oscillation in all magnetic-field directions for a generic model that describes a linked-trivial transition of a nodal ring and a nodal line, with the help of three self-consistent approaches, including the Fermi-surface tomography,  resistivity, and Landau fan diagrams. For the Hopf-link semimetal, we can always identify three distinct phase shifts, which evolve as the magnetic field is rotated, forming a unique pattern [Fig.~\ref{Fig:Model}(d)] that may facilitate identifying Hopf links in realistic materials, such as Li$_2$NaN (Fig.~\ref{Fig:Li2NaN}). 



\begin{figure}[tbp]
\centering
\includegraphics[width=0.48\textwidth]{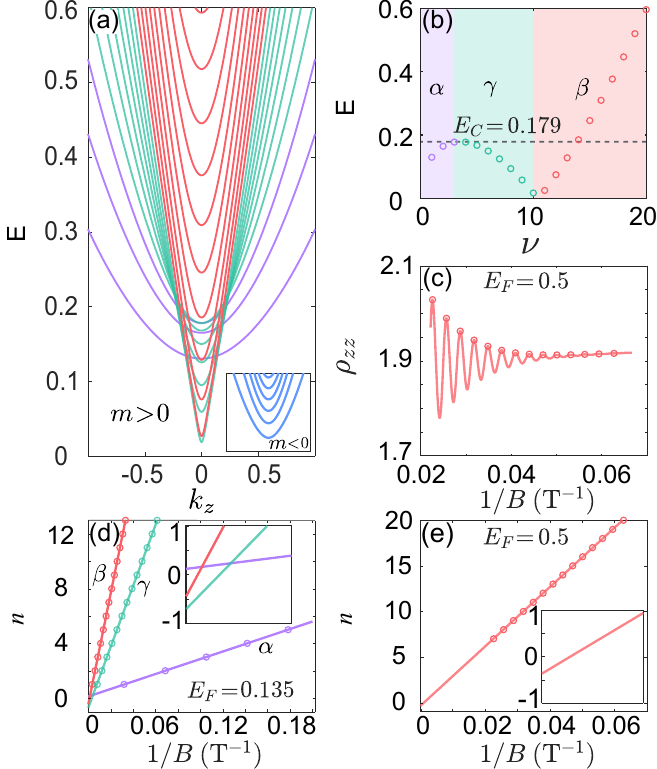}
\caption{Landau bands under a 30T $z$-direction magnetic field. (a) Energies of the Landau bands for the Hopf-link semimetal ($m=0.5$) as functions of the wave vector $k_z$. The purple, red, and green bands correspond to the $\alpha$, $\beta$, and $\gamma$ cross sections in Fig.~\ref{Fig:Model}(b), respectively. Inset: Landau bands for the trivial case ($m=-1$).
(b) The dependence of the band-bottom ($k_z=0$) energies on the Landau index $\nu$ distinguishes the $\alpha$, $\beta$, and $\gamma$ Landau bands. Above a critical energy $E_C=0.179$ (dashed line), only the $\beta$ bands contribute to band bottoms. 
(c) At the Fermi energy $E_F=0.5>E_C$,  the $z$-direction resistivity $\rho_{zz}$ as a function of $1/B$ contributed from the $\beta$ bands, where $B$ is the strength of the $z$-direction magnetic field. 
(d) The band index $n$ of the resistivity peaks from all bands as a function of $1/B$ at $E_F=0.135$.
Inset: Zoom-in of the intersects on the $n$ axis, which give the phase shifts listed in the $\phi = 0$ rows and Numeric $\Phi$ column in Tab.~\ref{Tab:phase}.
(e) The band index $n$ of the resistivity peaks from $\beta$ bands as a function of $1/B$ at $E_F=0.5$. Inset: Zoom-in of the intersect on the $n$ axis.}
\label{Fig:Landau}
\end{figure}

\textcolor{blue}{Phase shifts extracted from Fermi-surface tomography.}\label{S2} -- Figure \ref{Fig:Model}(a) shows the Fermi surface of a generic Hopf-link semimetal  ($m>0$), where the conduction and valence energy bands touch at some points of momentum (i.e., the nodes). The nodes form a nodal ring and a nodal line linked with each other. In a strong magnetic field, the Lorentz force forces electrons to perform the cyclotron motion, creating a confinement that quantizes the energy bands into a series of 1D bands of Landau levels (e.g., see Fig.~\ref{Fig:Landau}(a)). As the bands cross the Fermi energy one by one, the resistance or conductance may show a series of peaks, that is, the Shubnikov de Haas quantum oscillation. The quantum oscillation is characterized by its frequency $F$ and phase shift $\Phi$. The frequency $F=(\hbar/2\pi e)A$   
is determined by the Fermi-surface extremal cross-sectional area $A$ perpendicular to the magnetic field, i.e., the Onsager relation \cite{Onsager1952}, where $\hbar$ is the reduced Planck constant and $e$ is the elementary charge. In conventional metals, the phase shift $\Phi$ is believed to follow the rules \cite{lifshitz1956theory,Shoenberg84book}
\begin{equation}\label{Eq:Rules}
\Phi=-1 / 2+\Phi_B / 2 \pi+\Phi_{3 \mathrm{D}}, 
\end{equation}
where $\Phi_{B}$ is the Berry phase \cite{DiXiao2010RMP}, which is $\pi$ around a nodal line \cite{Mikitik99prl} due to the one-to-one correspondence between the momentum-dependent winding number and wavefunction of the
drumhead surface states \cite{Zhao23qf}. $\Phi_{3 \mathrm{D}} = \pm 1/8$ is a correction that occurs only in three dimensions.
For a maximal (minimal) cross section, $\Phi_{3 \mathrm{D}} = - 1/8$ (1/8) for electron carriers and $\Phi_{3 \mathrm{D}} = 1/8$ ($-1/8$) for hole carriers. 

However, the conventional rules have never been tested for the topological bands as sophisticated as those of Hopf-link semimetals. Therefore, we compare the phase shifts from the rules with those from a direct calculation of resistance, trying to give a signature of the Hopf link in the quantum oscillation measurement. Figure \ref{Fig:Model} (b) shows the extremal cross sections $\alpha$, $\beta$, and $\gamma$ in a magnetic field normal to the nodal-ring plane ($\mathbf{B}||z$). 
The three cross sections correspond to three oscillation frequencies, each has a phase shift. 
Throughout the work, we restrict our discussion for electron carriers. The outer maximal ($\gamma$) and inner minimal ($\beta$) cross sections along the nodal ring have no Berry phase, so their phase shifts are supposed to be $-1/2+0-1/8=-5/8$ and $-1/2+0+1/8=-3/8$.
The $\alpha$ minimal cross-section encloses the nodal line, thus has a $\pi$ Berry phase and a phase shift of $-1/2+\pi/2\pi+1/8=1/8$. Figure \ref{Fig:Model} (c) shows the $\delta$, $\zeta$, and $\eta$ extremal cross sections in a magnetic field parallel to the nodal-ring plane ($\mathbf{B}||x$). The minimal cross-section $\delta$ around the nodal ring has a $\pi$ Berry phase, so its phase shift is $-1/2+\pi/2\pi+1/8=1/8$.
The other two maximal cross sections $\eta$ and $\zeta$ have no Berry phase in the first Brillouin zone, resulting in the same phase shift $-1/2+0-1/8=-5/8$ for two different frequencies.
The phase shifts in other directions of magnetic field can be argued similarly, as summarized in the Analytic $\Phi$ column of Tab.~\ref{Tab:phase}.
In contrast, the inset of Fig.~\ref{Fig:Model}(a) shows that the trivial case ($m<0$) has only one Fermi surface. It has an only maximal cross section, regardless of the magnetic field's orientation.
It has a $\pi$ Berry phase and a phase shift of $-1/2+\pi/2\pi-1/8=-1/8$ in the $z$-direction magnetic field, and no Berry phase and a phase shift of $-1/2+0-1/8=-5/8$ in the $x$-direction magnetic field. The phase shifts of the only cross section in other directions of magnetic field are also summarized in the Analytic $\Phi$ column of Tab.~\ref{Tab:phase}.

Interestingly, Fig.~\ref{Fig:Model}(d) shows that, the summation of the phase shifts of all cross sections as a function of the angle of magnetic field $\phi$ form a unique pattern for the linked case (red shadow), in sharp contrast to the trivial case (blue shadow). This pattern can be understood as a poor man's conservation law of the phase shifts as the extremal cross sections transit with the direction of magnetic field. More importantly, the quantum oscillation measurement is readily accessible  \cite{Murakawa2016Science, WangCM2016PRL,LiCQ2018PRL, He2014PRL, Novak2015PRB, ZhaoYF2015PRX,Ratnadwip2017PNAS, Mazhar2016SciAdv, WangXF2016AEM, LvYY2016APL, HuJ2017PRB, PanHY2018SciR, HuJ2016PRL, HuJ2017PRB2, Kumar2017PRB}, so this pattern could serve as a signature to identify the Hopf link.

\textcolor{blue}{Phase shifts extracted from resistivity.} -- To verify the above results, we need to calculate the resistivity. We start with a generic Hamiltonian of Hopf-link semimetal
\begin{eqnarray}\label{Eq:Hamiltonian}
\mathcal{H}(\mathbf{k})=&& \left[k_x k_z+\left(m- Ck^2\right) k_y\right] \hat{\sigma}_x \nonumber\\
&&+ \left[-k_y k_z+\left(m-  Ck^2\right) k_x\right] \hat{\sigma}_y,
\end{eqnarray}
where $k^2=k_x^2+k_y^2+k_z^2$, $k_x,k_y,k_z$ are the wave vectors, $C=0.5$, $\hat{\sigma}_x$ and $\hat{\sigma}_y$ 
are the Pauli matrices \cite{YanZB2017PRB}.
When $m>0$, this model hosts a nodal line at $k_x=k_y=0$,as well as a nodal ring at $k_x^2+k_y^2=2m$ on the $k_z=0$ plane, forming a Hopf link, as shown in Fig.~\ref{Fig:Model}(a). 
When $m<0$, the model is trivial with no Hopf link, as shown in the inset of Fig.~\ref{Fig:Model}(a). 
We consider a magnetic field $\mathbf{B}=B(\sin\phi, 0, \cos\phi)$ that rotates in the $x$-$z$ plane, with the strength of magnetic field $B$ and the angle of magnetic field $\phi \equiv \arctan (B_x/B_z)$. We solve the 1D energy bands of Landau levels as functions of 
the wave vector $k_\parallel=k_x\sin{\phi} + k_z \cos{\phi}$ along the direction of magnetic field,
which remains invariant under the Lorentz force. 
When the magnetic field is along the $z$ direction ($\phi=0$), the eigen energies of the Landau bands can be solved analytically as 
\begin{eqnarray}
E_{k_z}^{\nu\pm}
=\pm \sqrt{2\nu [k_{z}^2+ M_{\nu}^2]}/\ell_B~,   
\end{eqnarray}
for the Landau index $\nu\ge 1$, and $E_{k_z}^0=0$
for $\nu=0$, where magnetic length $\ell_B\equiv\sqrt{1/e B}$, $M_{\nu} = m-k_z^2/2-(\nu+1/2)\omega_c$, and oscillation frequency $\omega_c \equiv 2C/\ell_B^2$.
The Planck constant $\hbar$ is set to 1 in this toy model. 
Figure \ref{Fig:Landau}(a) shows the Landau bands of the Hopf-link semimetal ($m=0.5$) when the magnetic field is along the $z$ direction. The bands consist of three sets (except the zero-energy $\nu=0$ band), as shown by the purple, red, and green bands,
corresponding to the $\alpha$, $\beta$, and $\gamma$ cross sections in Fig.~\ref{Fig:Model}(b), respectively. The energies of the $\alpha$ and $\beta$ bands from the minimal cross sections shift upward with increasing $\nu$. 
By contrast, those of the $\gamma$ bands from the maximal cross section shift downward with increasing $\nu$, as shown in terms of their band-bottom energies in Fig.~\ref{Fig:Landau}(b).
The inset of Fig.~\ref{Fig:Landau}(a) shows the Landau bands for the trivial case ($m=-1$), which consists of only one set, because there is only one maximal cross section, consistent with the inset of Fig.~\ref{Fig:Model}(a).

\begin{table}[htbp]
\centering
\begin{ruledtabular}
\centering
\caption{Phase shift $\Phi$ of the maximal (Max) or minimal (Min) cross sections for the electron Fermi pockets in Fig.~\ref{Fig:Model}, obtained by using Eq.~\eqref{Eq:Rules} (Analytic) and numerically calculated from the resistivity $\rho_{zz}$ 
or from the Landau fan diagram of band-bottom energies (Numeric). The angle of magnetic field is defined as $\phi \equiv \arctan (B_x/B_z)$. $\kappa$ is the maximal cross section of the trivial case.  
}\label{Tab:phase}
\begin{tabular}{ccccc}
\centering
$\phi$ & Cross  & Berry  &  Analytic  & Numeric  
\\
 & section &  phase &   $\Phi$ & $\Phi$ 
\\
\hline
& $\alpha$ (Min) &$\pi$ &1/8 & 0.115$\pm$0.016
\\
& $\beta$ (Min) &0&-3/8 & -0.366$\pm$0.011 ($\rho_{zz}$) \\
 0   & &  & & -0.390$\pm$0.007
\\
& $\gamma$ (Max) &0&-5/8 & -0.684$\pm$0.003
\\
& $\kappa$ (Max) & $\pi$ &-1/8 &  -0.141$\pm$0.009 ($\rho_{zz}$)
\\
&  &  & &-0.121$\pm$0.028
\\
\hline
& $\eta$ (Max) & $\pi$ & -1/8 & -0.129$\pm$0.000
\\
$\pi/8$& $\delta$ (Min)&$\pi$ &1/8 & 0.114$\pm$0.001
\\
& $\zeta$ (Max)&0 &-5/8 & -0.680$\pm$0.000
\\
& $\kappa$ (Max) & $\pi$ &-1/8 & -0.126$\pm$0.000
\\
\hline
& $\eta$ (Max) & $\pi$ &  -1/8 & -0.145$\pm$0.022
\\
$\pi/4$& $\delta$ (Min)&$\pi$ &1/8 & 0.146$\pm$0.015
\\
& $\zeta$ (Max)&0 &-5/8 & -0.561$\pm$0.007
\\
& $\kappa$ (Max) & $\pi$ &-1/8 & -0.149$\pm$0.001
\\
\hline
& $\eta$ (Max) & $\pi$ &  -1/8& -0.159$\pm$0.000
\\
$3\pi/8$& $\delta$ (Min)&$\pi$ & 1/8 & 0.099$\pm$0.002
\\
& $\zeta$ (Max)&0 & -5/8 & -0.584$\pm$0.008
\\
& $\kappa$ (Max) & $\pi$ & -1/8 & -0.168$\pm$0.000
\\
\hline
&$\eta$ (Max)&0 &-5/8 & -0.650$\pm$0.000
\\
$\pi/2$ &$\delta$ (Min)&$\pi$ &1/8 & 0.127$\pm$0.000
\\
&$\zeta$ (Max)&0 &-5/8 & -0.624$\pm$0.026
\\
& $\kappa$ (Max) & 0 &-5/8 & -0.677$\pm$0.000
\\
\end{tabular}
\end{ruledtabular}
\end{table}

To capture the phase shift, we calculate the resistivity $\rho_{zz}=1/\sigma_{zz}$ along the $z$ direction of the Landau bands 
(details in Sec.~SIII of \cite{Supp}), with the conductivity
\begin{equation}
\sigma_{z z}=\frac{e^2}{2 \pi \ell_B^2} \sum_{n\in \nu,\pm}  \int \frac{d k_z}{2 \pi} \tau_{k_x, k_z}^{n}\left(v_z^{n, k_z}\right)^2 \delta\left(E_{k_z}^{n}-E_F\right)~.
\end{equation}
The $z$-direction velocity of band $n$ at $k_z$ for degenerate states is calculated by
$v_z^{n,k_z}=\partial E_{k_z}^{n} /  \partial k_z$, and
$\tau_{k_x, k_z}^{n}$ is the scattering time of band $n$ at $k_z$ with the degenerate Landau levels denoted by $k_x$.
Due to the sophisticated expression of the scattering time, we calculate the conductivity numerically. 
When the Fermi level $E_F$ is set to 0.5, only the $\beta$ band bottoms can cross the Fermi energy, as shown in Fig.~\ref{Fig:Landau}(b).
The corresponding $\rho_{zz}$ as a function of $1/B$ with a series of peaks is shown in Fig.~\ref{Fig:Landau}(c). 
The phase shift can be captured by searching the intersect of the band index $n$ as a function of $1/B$ at the peak positions.
We find $\Phi=-0.366\pm 0.011$ for the $\beta$ cross section at $\phi=0$, consistent with analytic $\Phi$ = $-3 /8$ in Tab. \ref{Tab:phase}.
For the trivial case, the phase shift extracted from $\rho_{zz}$ is $-0.141\pm 0.009$ (Fig.~S1 of \cite{Supp}), close to $-1/8$ in Tab. \ref{Tab:phase} for the $\kappa$ cross section at $\phi=0$.

\textcolor{blue}{Phase shifts extracted from Landau fan diagram of band-bottom energies} -- The above results infer that the resistivity peaks correspond to the bottoms of the Landau bands.
Specifically, the $n$-$1/B$ relation is found at those values of $B$ when the bottom of the $n$-th band crosses the Fermi energy, i.e., Landau Fan diagram. 
In Fig.~\ref{Fig:Landau} (c), there is only one frequency from the $\beta$ cross section, because the Fermi energy $E_F$ is higher than the critical value $E_C$. When $E_F<E_C$, 
Fig.~\ref{Fig:Landau} (d) shows
three $n$-$1/B$ plots with different slopes, corresponding to all three cross sections in Fig.~\ref{Fig:Model} (b).
The phase shifts extracted from Fig.~\ref{Fig:Landau} (d) are 
listed in the Numeric column in Tab. \ref{Tab:phase} at $\phi= 0$, which are consistent with those in the Analytic column. It can further justifying the effectiveness of our approach method to capture the phase shifts. 
Therefore, we generalize the approach to arbitrary directions of magnetic field by numerically calculating the Landau bands (details in Sec.~SIV of \cite{Supp}), as shown by the Numeric column in Tab.~\ref{Tab:phase}, for $\phi=\pi/8$, $\pi/4$, $3\pi/8$, and $\pi/2$, where $\phi$=$\pi/2$ corresponds to $\mathbf{B}||x$.

\textcolor{blue}{Realistic material Li$_2$NaN} - 
Now we apply the above approach to a realistic material Li$_2$NaN \cite{Lenggenhager21prbl}. 
The energy spectrum near the Fermi energy of Li$_2$NaN can be described by a four-band tight-binding model 
\begin{equation}\label{Eq:LiNaN}
\mathcal{H}_{\rm eff}=\sum_{i\alpha}\varepsilon_{\alpha}c^{\dagger}_{i\alpha}c_{i\alpha}+\sum_{i\alpha,j\beta}
t^{i,j}_{\alpha,\beta}
c^{\dagger}_{i\alpha}c_{j\beta},
\end{equation}
where $c^{\dagger}_{i\alpha}$ ($c_{i\alpha}$) are electron creation (annihilation) operators on the $\alpha$ orbital at lattice site $i$, $\varepsilon_{\alpha}$ is the on-site energy of the $\alpha$ orbital, and $t^{i,j}_{\alpha,\beta}$ is the hopping integral between the $\alpha$ orbital on site $i$ and $\beta$ orbital on site $j$ (detailed expressions in Sec.~SV of \cite{Supp}). 
The Hopf link described by this model is shown by the blue part in Fig.~\ref{Fig:Li2NaN}(b). 
The red part in Fig.~\ref{Fig:Li2NaN}(b) represents some additional nodal lines well separated from the nodal line of the Hopf-link in both momentum and energy. Their effects can be excluded by choosing proper chemical potentials when analyzing the phase shifts. 

 
\begin{table}[htbp]
\centering
\begin{ruledtabular}
\centering
\caption{Numerically calculated phase shifts from Landau fan diagrams for Li$_2$NaN.
The angle of magnetic field is defined as $\phi\equiv\arctan (B_x/B_z)$. 
The $\alpha$, $\beta$, $\gamma$, $\eta$, $\delta$ and $\zeta$ indicate the corresponding cross sections in Tab. \ref{Tab:phase}.}\label{Tab:Li2NaN}
\begin{tabular}{ccccc}
\centering
$\phi$ &  & Phase shift $\Phi$ &   
\\
\hline
$0$  &0.115$\pm$0.000 ($\alpha$) &-0.372$\pm$0.016 ($\beta$) &-0.601$\pm$0.001 ($\gamma$) 
\\
\hline
$\pi/8$  &-0.119$\pm$0.004 ($\eta$) &0.111$\pm$0.006 ($\delta$) &-0.616$\pm$0.029 ($\zeta$) 
\\
\hline
$\pi/4$  &-0.164$\pm$0.000 ($\eta$) &0.153$\pm$0.001 ($\delta$) &-0.595$\pm$0.001 ($\zeta$) 
\\
\hline
$3\pi/8$  & -0.180$\pm$0.000 ($\eta$) & 0.161$\pm$0.009 ($\delta$) & -0.587$\pm$0.000 ($\zeta$) 
\\
\hline
$\pi/2$  & -0.571$\pm$0.000 ($\eta$) & 0.118$\pm$0.002 ($\delta$) & -0.624$\pm$0.053 ($\zeta$) 
\end{tabular}
\end{ruledtabular}
\end{table}

By applying the magnetic field $\mathbf{B}=B(\sin\phi, 0, \cos\phi)$ to the tight-binding model, 
the hopping integrals become carrying the flux of magnetic field in terms of a path integral
$t^{i,j}_{\alpha,\beta}
\rightarrow t^{i,j}_{\alpha,\beta} \exp [i (2 \pi/\phi_0) \int_{R_j}^{R_i}  d \boldsymbol{\ell}  \cdot \mathbf{A}]$,  
where the flux quantum $\phi_0=h/e$ and $\mathbf{A}$ is the vector potential of the magnetic field (Sec.~SV of \cite{Supp}). 
The Landau bands of Li$_2$NaN, as shown in 
Fig.~\ref{Fig:Li2NaN}(c) is calculated by diagonalizing the effective tight-binding model under a $z$-direction magnetic field.
The Landau bands also can be identified by three sets. We use the band-bottom energies [dashed boxes in Fig.~\ref{Fig:Li2NaN}(d)] to plot the $n$-$1/B$ relation in Fig.~\ref{Fig:Li2NaN}(e), to extract the three phase shifts 0.115, -0.372, and -0.601, respectively, as listed in Tab.~\ref{Tab:Li2NaN} at $\phi = 0$. The results are consistent with the phase shifts found from the generic model in Tab.~\ref{Tab:phase} for the Analytic and Numeric columns at $\phi=0$. The phase shifts in other directions of magnetic field can be found similarly (Sec. {SV} of \cite{Supp}), as shown in Tab.~\ref{Tab:Li2NaN} for $\phi=\pi/8$, $\pi/4$, $3\pi/8$, and $\pi/2$, which are also consistent with those from the generic model in Tab. \ref{Tab:phase}.
More importantly, the phase shifts of Li$_2$NaN also form a similar pattern as that of the generic model, as compared by the green and red circles in Fig.~\ref{Fig:Model}(d).
The calculation indicates that a magnetic field of around 10 T is needed to extract the phase shifts of Li$_2$NaN.

\textcolor{blue}{Discussions} -- The consistence between the numerically calculated phase shifts and analytic rule in Eq.~\eqref{Eq:Rules} shows that the phase shift is mainly determined by the topology, rather than subtle details, of the Fermi surface. Weyl semimetals do not have the torus Fermi surface with multiple frequencies and Normal nodal-line semimetals do not have an $\eta$ extremal cross section, so they can be distinguished from Hopf-link semimetals.

\begin{acknowledgements}
{\color{blue}Acknowledgments} -- $^\dag$Lei Shi, Xiaoxiong Liu, and C. M. Wang contribute equally. We thank helpful discussions with Xiaoqun Wang and Xiangang Wan. This work was supported by the National Key R\&D Program of China (2022YFA1403700), Innovation Program for Quantum Science and Technology (2021ZD0302400), the National Natural Science Foundation of China (12350402, 12304196, 12474048, and 11925402), Guangdong Basic and Applied Basic Research Foundation (2022A1515111034 and 2023B0303000011), Guangdong province (2020KCXTD001), Guangdong Provincial Quantum Science Strategic Initiative (GDZX2201001 and GDZX2401001), the Science, Technology and Innovation Commission of Shenzhen Municipality (ZDSYS20190902092905285), Center for Computational Science and Engineering of SUSTech, and the New Cornerstone Science Foundation for the support through the XPLORER PRIZE. 
\end{acknowledgements}

\bibliographystyle{apsrev4-2-etal-title}
\bibliography{refs-transport,Quantum_Oscillation}
\end{document}